# Distributed Bragg reflector-mediated excitation of InAs/InP quantum dots emitting in the telecom C-band

A. Musiał[1,a)], M. Wasiluk[1], M. Gawełczyk[2], J. P. Reithmaier[3], M. Benyoucef [3,a)], G. Sęk[1] and W. Rudno-Rudziński[1]

[1]*Department of Experimental Physics, Faculty of Fundamental Problems of Technology, Wrocław University of Science and Technology, Wybrzeże Wyspiańskiego 27, 50-370 Wrocław, Poland*

[2] *Department of Theoretical Physics, Faculty of Fundamental Problems of Technology, Wrocław University of Science and Technology, Wybrzeże Wyspiańskiego 27, 50-370 Wrocław, Poland*

[3]*Institute of Nanostructure Technologies and Analytics (INA), Center for Interdisciplinary Nanostructure Science and Technology (CINSaT), University of Kassel, Heinrich-Plett-Str. 40, 34132 Kassel, Germany*

Anna Musiał, anna.musial@pwr.edu.pl, ORCID: 0000-0001-9602-8929

Maja Wasiluk, maja.wasiluk@pwr.edu.pl, ORCID: 0000-0002-6657-3827

Michał Gawełczyk, michal.gawelczyk@pwr.edu.pl, ORCID: 0000-0003-2299-140X

Johann Peter Reithmaier, jpreith@physik.uni-kassel.de, ORCID: 0000-0002-1974-8292

Mohamed Benyoucef, m.benyoucef@physik.uni-kassel.de, ORCID: 0000-0002-4756-3818

Grzegorz Sęk, grzegorz.sek@pwr.edu.pl, ORCID: 0000-0001-7645-8243

Wojciech Rudno-Rudziński, wojciech.rudno-rudzinski@pwr.edu.pl, ORCID: 0000-0002-3475-7517

---

a) Corresponding author: anna.musial@pwr.edu.pl, m.benyoucef@physik.uni-kassel.de

\
Table of contents

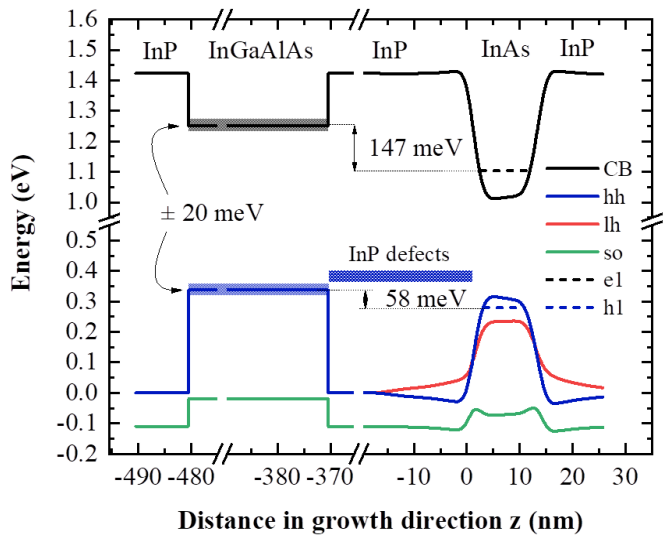

We show that states in the InGaAlAs layer of the distributed Bragg reflector can take part in carrier relaxation to InAs/InP quantum dots (QDs). This mechanism is efficient up to approx. 100 K and depends on the local QD environment. Described transfer of holes has important consequences for the design and performance of QD-based devices.

# Distributed Bragg reflector-mediated excitation of InAs/InP quantum dots emitting in the telecom C-band

A. Musiał[1,a)], M. Wasiluk[1], M. Gawełczyk[2], J. P. Reithmaier[3], M. Benyoucef [3,a)], G. Sęk[1] and W. Rudno-Rudziński[1]

[1]*Department of Experimental Physics, Faculty of Fundamental Problems of Technology, Wrocław University of Science and Technology, Wybrzeże Wyspiańskiego 27, 50-370 Wrocław, Poland*

[2] *Department of Theoretical Physics, Faculty of Fundamental Problems of Technology, Wrocław University of Science and Technology, Wybrzeże Wyspiańskiego 27, 50-370 Wrocław, Poland*

[3]*Institute of Nanostructure Technologies and Analytics (INA), Center for Interdisciplinary Nanostructure Science and Technology (CINSaT), University of Kassel, Heinrich-Plett-Str. 40, 34132 Kassel, Germany*

**Abstract**

We demonstrate that optical excitation of InAs quantum dots (QDs) embedded directly in an InP matrix can be mediated via states in a quaternary compound constituting an InP/InGaAlAs bottom distributed Bragg reflector (DBR) and native defects in the InP matrix. It does not only change the carrier relaxation in the structure but could also lead to the imbalanced occupation of QDs with charge carriers, because the band structure favors the transfer of holes. Thermal activation of carrier transfer can be observed as an increase in the emission intensity versus temperature for excitation powers below saturation on the level of both an inhomogeneously broadened QD ensemble and single QD transitions. That increase in the QD emission is accompanied by a decrease in the emission from the InGaAlAs layer at low temperatures. Finally, carrier transfer between the InGaAlAs layer of the DBR and the InAs/InP QDs is directly proven by the photoluminescence excitation spectrum of the QD ensemble. The reported carrier transfer can increase the relaxation time of carriers into the

---

b) Corresponding author: anna.musial@pwr.edu.pl, m.benyoucef@physik.uni-kassel.de

QDs and thus be detrimental to the coherence properties of single and entangled photons. It is important to take it into account while designing QD-based devices.

**Introduction**

For exploiting the optical properties of semiconductor heterostructures, it is essential to understand the relaxation processes leading to the occupation of the lowest-energy state from which emission takes place. Corresponding relaxation time constitutes a fundamental limitation for the modulation speed of lasers and optical amplifiers. In the case of quantum devices, carrier relaxation also plays a crucial role. For instance, the existence of many relaxation paths introduces timing jitter in the emission of single photons and reduces their degree of indistinguishability [Hub2015]. In the case of qubits, carrier dynamics limits the maximal rate of logic operations that can be performed. In both examples, all long-lived states that can capture carriers during relaxation and release them on a long-time scale after the excitation pulse are disadvantageous. As exemplified, it is important to determine the carrier dynamics and efficient ways of populating the QD states, as well as to identify processes hindering it for newly developed materials or physical systems. Understanding the possible excitation schemes and excitation transfer processes reveals the means of controlling them and therefore opens up engineering possibilities.

It is especially important for operation at the telecommunication wavelengths, which is desirable for many applications. Particularly appealing is the telecom C-band – 1530-1565 nm – providing the lowest optical signal loss in silica fibers (0.15 dB/km) and, therefore, the largest possible range of data transmission without amplification. Obtaining emission in this spectral range is straightforward when InP-substrate-based heterostructures are used. InAs QD ensembles have been widely employed in optoelectronics as active regions of lasers and optical amplifiers [Rei2007]. Using single InP-technology-based nanostructures allowed demonstrating single-photon emission under electrical [Miy2008] and optical excitation [Ara2020, Hol2022] with probabilities of multiphoton events down to $4.4 \times 10^{-4}$ [Miy2016]. Despite the abovementioned achievements with InAs/InP QD structures and proof of concept demonstration of their application potential [And2020, Tak2015, Mül2018], there are still aspects that need to be addressed. For instance, the structural quality and photon extraction efficiency [Mus2021]. The alternative approach to achieve emission in the telecom C-band is to use a metamorphic buffer layer for the GaAs substrate [Por2019]. These quantum emitters have been proven high-purity sources of single indistinguishable photons [Naw2019] and high-fidelity sources of entangled photon pairs [Lett2021, Zeu2021].

In this work, we investigate symmetric InAs QDs embedded directly in the InP matrix, with the potential for generating polarization-entangled photon pairs from a biexciton-exciton cascade [Kor2018]. We focus on the influence of an InGaAlAs layer within the bottom distributed Bragg reflector (DBR) on the emission from the InAs/InP QDs. At first, thermally activated excitation power-dependent increase in emission intensity is shown for the QD ensemble for temperatures in the range up to 100 K. Carrier transfer from the InGaAlAs layer to the QD states is further directly proven by a photoluminescence excitation (PLE) experiment. A similar effect is also observed on the level of single QD transitions. Eventually, the influence of the observed DBR-mediated QD excitation on the performance of quantum devices is discussed.

**Experimental**

The QD structure investigated in this work consists of molecular beam epitaxy (MBE) grown InAs QDs in InP matrix on an Fe-doped InP (100) substrate. A single QD layer is grown at 490°C on top of an InP/In$_{0.53}$Ga$_{0.37}$Al$_{0.1}$As DBR (25 pairs), followed by a 246 nm thick InP layer, and it is capped with an InP layer of the same nominal thickness (246 nm). The QDs feature low spatial density of $5 \times 10^8$ cm$^{-2}$ to $2 \times 10^9$ cm$^{-2}$ and high in-plane symmetry indicated by a low fine-structure splitting of neutral exciton states [Kor2018]. These characteristics are obtained due to an additional ripening step during epitaxial growth [Ben2013, Yac2014]. What distinguishes the investigated structure from standard QDs in the strong confinement regime is the relatively large volume [Rud2021, Kor2018] (a base outer diameter of the embedded QDs in the range of 55 ± 15 nm – based on the cross-sectional transmission electron microscopy - TEM - image not shown here [Car2017]) and weaker quantization in the growth direction (QD heights up to 15 nm [Rud2021]). The InGaAlAs material is lattice-matched to InP and therefore enables strain-free growth of a thick (almost 6 μm) DBR structure consisting of 25 pairs of layers [Kor2018]. Its energy bandgap spans a broad range, depending on the exact alloy composition [Gra2010, Mon1992]. So far, this quaternary material has been mainly used as a barrier and waveguide material for laser structures, typically with 53% of In and 23% of Ga. In our case, however, it constitutes one of the DBR sections. Therefore, a different material composition - the one maximizing refractive index contrast, with 37% of Ga and 53% of In, is selected. This results in a bandgap narrower in comparison to 1.1 eV for more commonly used material composition (the room temperature value). Therefore, the corresponding band edges are energetically much closer to the QD states. We present results from two sample pieces from the same wafer – a planar structure (study of the QD ensemble) and a patterned structure with micrometer-sized cylindrical mesas etched on the sample surface (single QD emission). For mesa fabrication, dry etching by inductively coupled plasma reactive ions has been used after writing a regular pattern with electron-beam lithography using an $SiO_2$ mask.

In the case of temperature-dependent experiments on the QD ensemble, a macro photoluminescence (PL) setup with a standard lens (10 cm focal length, resulting in approx. 100 μm diameter of the laser spot on the sample surface) is used. The planar sample (without mesas) is mounted in a He refrigerator, allowing temperature control from room temperature down to 10 K. Optical signal is dispersed by a 0.3-m-focal length monochromator before being directed to an InGaAs multichannel linear array detector. Non-resonant excitation with a 640 nm semiconductor laser is used for all the measurements.

Photoluminescence excitation spectra are collected with similar spectral resolution (0.32-m-focal length monochromator) but with a higher spatial resolution (2 μm diameter spot) and excitation density provided by a microscope objective with a 0.4 numerical aperture. A combination of Ti:Sapphire laser (3 ps pulses at 76 MHz repetition rate) with an optical parametric oscillator is used as an excitation source, providing a broad range of tuneability of the excitation energy.

For single QD experiments, a microphotoluminescence setup with high spectral (20 μeV) and spatial (2 μm) resolution optimized for the telecom spectral range is used. A long working distance (20 mm) infinity-corrected microscope objective with a numerical aperture of 0.4 focuses the excitation beam on the sample surface and collects the QD emission. The detection system consists of a 1 m-focal length monochromator and a multichannel InGaAs linear array photodetector. The sample is mounted on a cold finger of a liquid He flow cryostat with a temperature sensor and a heater enabling temperature-dependent measurements in the range of 4.5 – 300 K.

**Theoretical modeling**

We theoretically model the examined structures to support our considerations and interpretation of the experimental results. For this purpose, we use the available structural data from cross-sectional TEM scans, indicating that the QDs have a lens shape with a base diameter of about 55 nm and a height that can reach 10-15 nm (we assume 12 nm in the calculations) [Car2017]. The TEM image shows that the distribution of material inside the dot is inhomogeneous, with a greater concentration of arsenic in the center reaching 80%. To model this and to fine-tune the ground state energy to the observed emission, we assume pure InP bulk material as the barrier, 85% As concentration in the QD center, and 42% in the outer part of the dot and in the 1.2 nm thick wetting layer. The material distribution modeled in this way is subjected to Gaussian averaging to simulate the interdiffusion of atoms at the interface.

To calculate the eigenstates of electrons and holes in such a model quantum dot, we use the implementation [Gaw2014] of the multiband envelope-function $\boldsymbol{k}\cdot\boldsymbol{p}$ method [Bah1990]. The calculation includes the structural strain computed within the theory of continuous elasticity, the shear strain-induced piezoelectric field with terms up to the second order in the strain-tensor components, and spin-orbital effects resulting from the lack of bulk inversion symmetry of zincblende materials. The exact form of the $\boldsymbol{k}\cdot\boldsymbol{p}$ Hamiltonian and details of implementation can be found in Refs. Mie2018, Gaw2018, and the material parameters used by us in Refs. Hol2020, Gaw2017 (and references therein). By calculating energies of the few lowest electron and hole states within the QD and the band edges of respective subbands, we can estimate the characteristic energies for possible thermally activated processes.

We use the same tools and methods for the InGaAlAs layer of the bottom DBR structure. In this case, however, we neglect interface effects and consider the layer as ideal and in-plane infinite. The mesoscopic thickness of the layer justifies such treatment.

**Results and discussion**

First, we discuss the results obtained for the QD ensemble. In Fig. 1a), a low-temperature PL spectrum measured for intermediate excitation power is presented in a logarithmic scale. For these excitation conditions, emission from both the QD ensemble (centered at 0.818 eV) and the InGaAlAs layer (0.914 eV) is observed. For lower excitation, emission from the QD ensemble dominates, while emission from the InGaAlAs layer takes over for higher excitation. The emission energy of the InGaAlAs layer confirms that the Al content in the InGaAlAs layer must be lower than 12%, and Ga content is close to the nominal 37%, together with 51% of In [Gra2010, Mon1992].

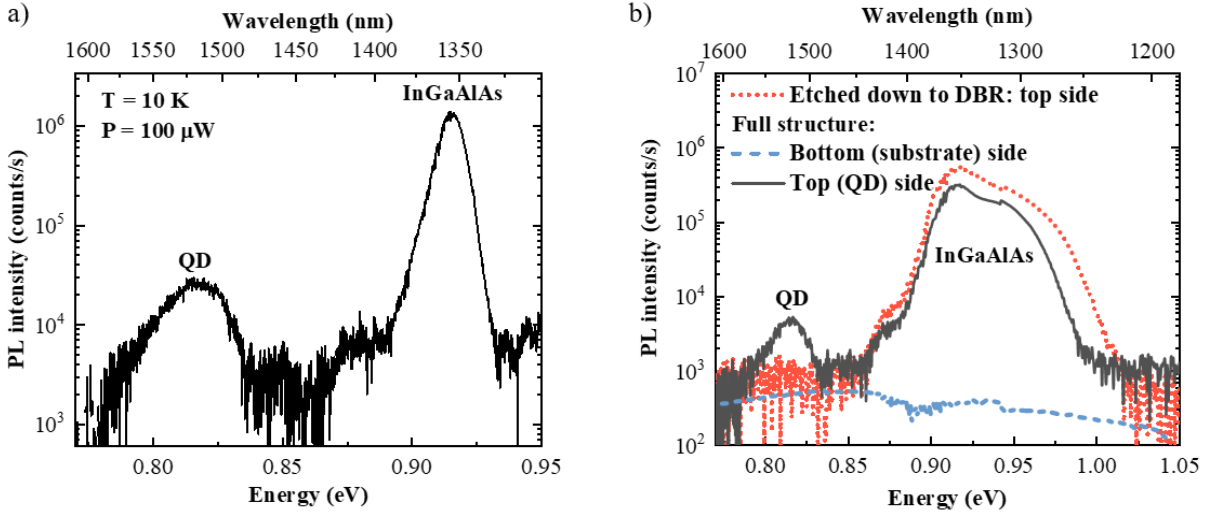

**Figure 1.** a) Low-temperature PL spectrum measured at intermediate excitation conditions; b) Low-temperature PL spectra for high excitation measured for structure etched down to the DBR (dotted red line) and planar sample in a standard PL configuration (solid black line) and for structure excited and signal collected from the substrate side (dashed blue line).

Additionally, we confirm that the emission centered at 0.914 eV originates spatially from the DBR section. We achieve this by comparing emission from the top and bottom of the complete structure with the emission from the top side of the structure etched down to the DBR section (Fig. 1b)). These measurements were performed at low temperature and high excitation conditions. In the case of emission from the top of the complete structure (a standard experimental configuration), a PL spectrum similar to the one presented in Fig. 1a) is observed. When the same structure is excited (and signal collected) from the bottom of the sample (the substrate side), no spectral features are observed in the spectral range of interest. The distance from the QD layer is too large to excite them, and the lack of high energy emission suggests that it is not related to the InP material (defect states from the InP matrix could emit in this spectral range [Gra2005, Wil1990]). Finally, the patterned structure was investigated. The etching depth of 600 nm ensures that in the region outside mesas, only the DBR on the InP substrate is left. In this emission spectrum, the high-energy emission persists.

As the InP-related origin of this emission was previously excluded, we are left with the InGaAlAs layer (including interfaces and volume defect states). The emission energy agrees with the expected bandgap energy of InGaAlAs, and the cross-sectional TEM image shows smooth interfaces within the DBR section (not shown here). The high optical quality of the DBR is also reflected in high stopband reflectivity, exceeding 98% [Kor2018]. Therefore, we conclude that the high-energy emission comes from the InGaAlAs layer within the DBR and corresponds to its bandgap energy.

In the next step, we verify whether InGaAlAs states influence the thermal stability of emission from the QD ensemble. We investigate the temperature dependence of PL intensity in a broad temperature range, from room temperature down to 10 K. In Fig. 2a) the integrated QD PL intensity as a function of temperature is presented for three excitation powers. For all cases, an initial emission intensity increase is observed (Table 1 and Fig. 2b)). The amplitude of the intensity increase decreases with excitation power. This is also true for the temperature at which maximum of the intensity is reached.

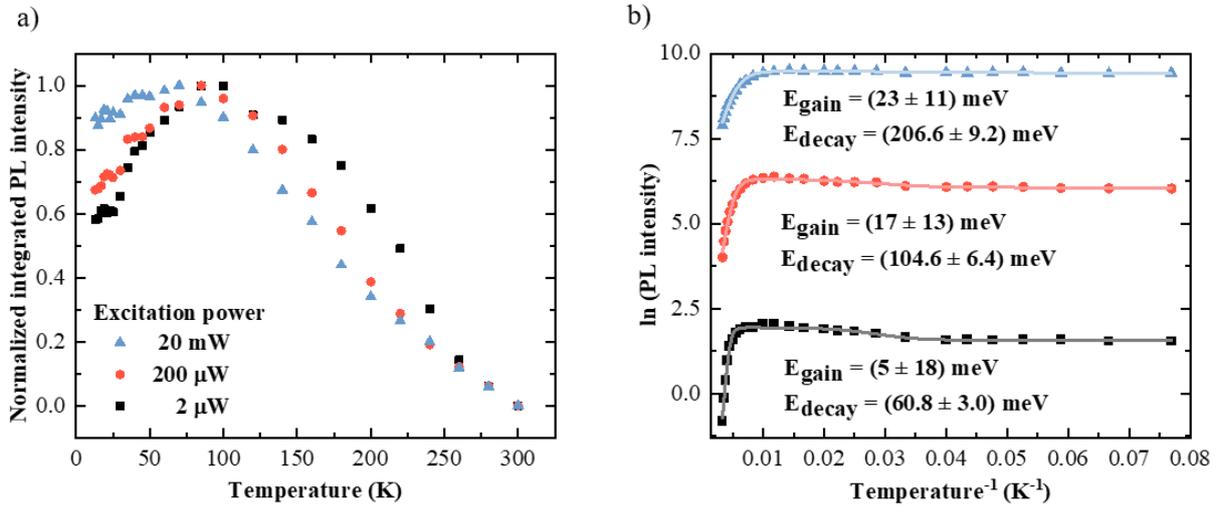

**Figure 2.** Temperature dependence of PL intensity integrated over the spectral range of QD ensemble emission presented in a) linear scale and b) Arrhenius analysis (solid lines) for three excitation powers: $P = 2$ μW (black squares), $10^2$ P (red circles) and $10^4$ P (blue triangles) – maximal achievable in PL configuration.

**Table 1.** Summary of parameters determined experimentally from quantitative analysis of the temperature-dependent PL of QD ensemble.

| Excitation power | Relative increase of the PL intensity | Temperature at which PL intensity is maximal | Activation energy (PL intensity | Activation energy (PL intensity |
|---|---|---|---|---|

| (P = 2 μW) | | | increase) | quenching) |
|---|---|---|---|---|
| P | 1.70 | 100 K | 5 ± 18 meV | 206.6 ± 9.2 meV |
| $10^2$ P | 1.50 | 85 K | 17 ± 13 meV | 104.6 ± 6.4 meV |
| $10^4$ P | 1.15 | 70 K | 23 ± 11 meV | 60.8 ± 3.0 meV |

The increase in the PL intensity cannot be caused by carrier redistribution within the QD ensemble, as we are integrating over the whole spectral range in which the QD emission can be seen. Therefore, the observed behavior indicates a thermally activated transfer of carriers from other states in the structure into the QDs. In the case of non-resonant excitation, the carriers are absorbed mainly in the barrier material (InP), and then they relax to the lowest energy state in the structure (QDs). The excess energy is typically dissipated into the crystal lattice (phonon relaxation), which occurs on the picosecond timescale [Mis2008]. Because this is much shorter than the typical radiative recombination times in the structure, the emission is observed only from the structure's ground state. Observation of emission from higher-energy states in the structure requires a state-filling effect and can be achieved by increasing excitation power. We will refer to this further as direct (optical) excitation.

Another scenario can be realized if there are local minima of potential within the structure. In that case, during the relaxation process, the carriers can be captured in these minima before they reach the lowest energy state (the global minimum – QD states) [Olb2017, Hol2020b]. Suppose the thermal energy at a given temperature is smaller than the localization energy. The carrier is then trapped in the intermediate state and cannot further relax to the global minimum of energy. Thus, it recombines from the state it was captured in. Carriers can be released from such a state if the temperature is increased (thermally activated process) or if tunneling to other states in the structure is possible. In that case, they can be transferred to the lowest energy QD states and add up to its emission. However, one has to keep in mind that if the QD states are already fully occupied from the direct optical excitation (high excitation conditions), the carriers released thermally from the confining potential minima will not contribute to the QD emission intensity anymore. As a result, no increase in the QD PL intensity as a function of temperature will be observed. Additional carriers can be supplied to the QDs only when they are not fully occupied at a given moment by the direct optical excitation.

In the case of continuous excitation, at a given excitation power and temperature, a certain probability can be assigned to each of the processes: trapping carriers in the intermediate state, transferring carriers to the QD states, and direct occupation of the QD states. The balance between these processes is reflected in the temperature at which the maximum of the QD emission intensity occurs and depends on their characteristic energy scales (depth of confining potential for the intermediate and QD states). The observed trends in the relative changes of emission intensity and in temperature, at which the maximal PL intensity is observed (Table 1), can be explained using this scenario. First, one has to note that the total density of states in the QDs is finite and significantly lower than in bulk. Thus, their emission can be saturated when the number of available carriers (proportional to the excitation power) and their capture rate into the QDs are high enough. Hence, the QD states are instantaneously reoccupied after they emit, and the dot is fully occupied all the time. It is reflected in the saturation of the emission intensity as a function of excitation power. Indeed, excitation power-dependent measurements at 15 K (Fig. 3) reveal deviation from the linear increase in the PL intensity for excitation powers exceeding 2 mW ($10^3$P).

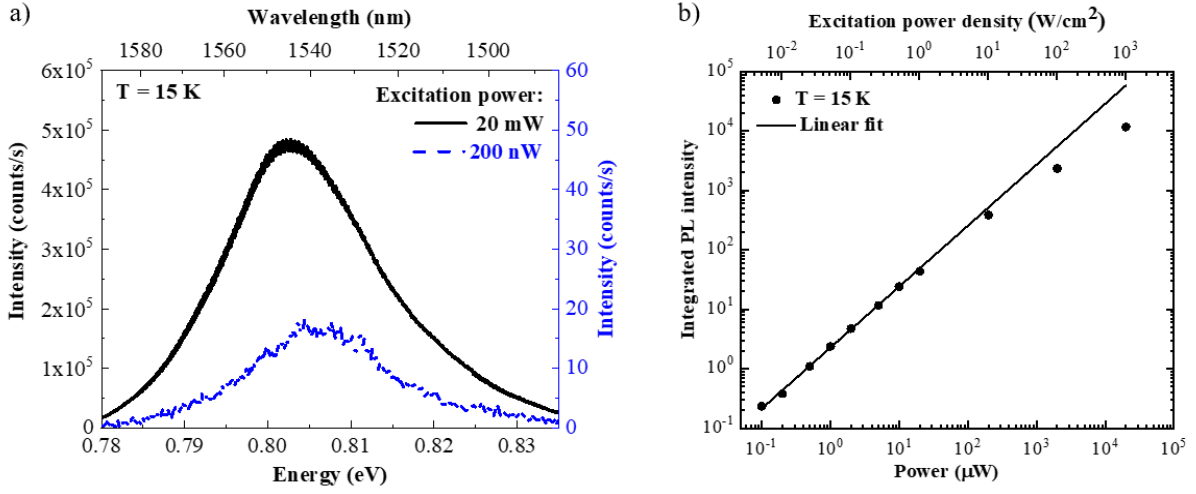

**Figure 3.** a) Exemplary low-temperature PL spectra for low excitation power - 200 nW (dotted blue line) and high excitation power - 20 mW (solid black line); b) dependence of integrated PL intensity of QD ensemble (symbols) at low temperature ($T = 15$ K) with a linear fit to low excitation range of experimental data (solid line).

For higher excitation powers, more QD states are already directly occupied at low temperature, and there are fewer unoccupied states that can be populated via additional transfer of carriers into the QDs. As a result, the amplitude of the PL intensity increase with temperature is lower, and the saturation conditions are reached for lower temperature.

To describe the obtained results quantitatively, we determined the energies that could be associated with both the increase and decrease in the PL intensity from the QD ensemble using the empirical modified Arrhenius function after Ref. Olb2017:

$$I(T) = \frac{I(0) + \frac{I_{\text{gain}}}{1 + C_{\text{gain}}\exp\left(-\frac{E_{\text{gain}}}{kT}\right)}}{1 + C_{\text{decay}}\exp\left(-\frac{E_{\text{decay}}}{kT}\right)},$$

where $I(T)$ is the temperature-dependent emission intensity, $I_{\text{gain}}$ corresponds to the intensity gained due to the transfer of carriers to the QDs, $E_{\text{gain}}$ and $E_{\text{decay}}$ are the activation energies, $k$ is the Boltzmann constant, $C_{\text{gain}}$ and $C_{\text{decay}}$ are efficiencies of the respective processes. These results are indicated in Fig. 2b) and summarized in Table 1. Arrhenius analysis of the temperature dependence of the emission intensity revealed activation energies for PL intensity increase below 25 meV and activation energies related to the PL quenching mechanisms in the range of 60, 100, and 200 meV, and decreasing with increasing excitation power. To interpret the experimentally determined energies, one has to look at the calculated energy band structure shown in Fig. 4.

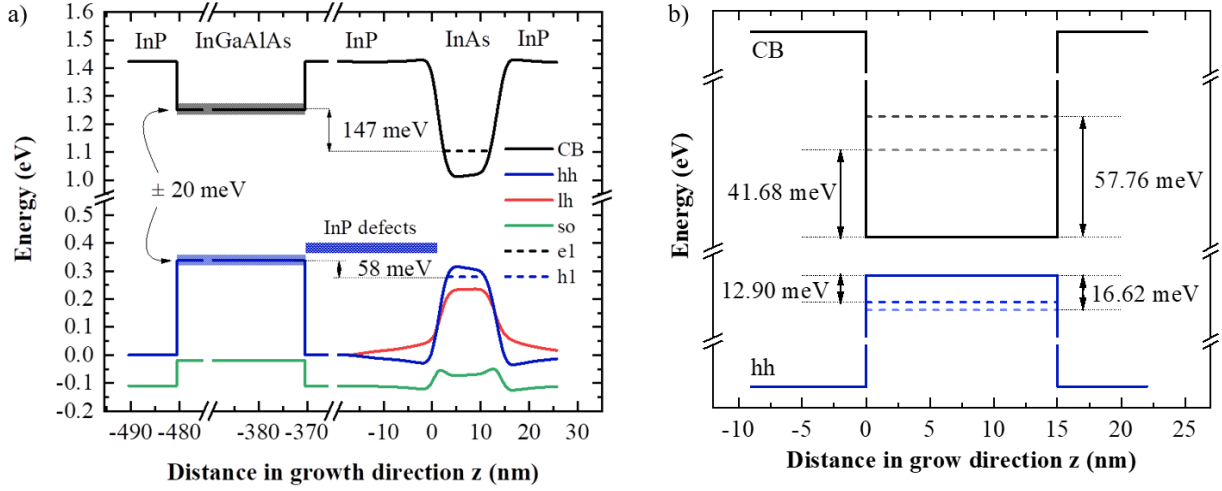

**Figure 4.** a) Conduction (black) and valence (blue for heavy holes – hh, red for light holes – lh and green for spin-orbit split-off subband) band edges for nominal structure parameters together with the range of energies characteristic for native InP defects (see text for more details); b) first two electron (black) and heavy hole (blue) states confined in the typical QD presented with idealized band-edge profiles. The calculations are done for a QD with 55 nm outer base diameter, 12 nm height and P admixture varying from 15% in the QD center to 38% in its outer layer.

The first important observation is that the valence band edge of the InGaAlAs and InAsP material almost coincide in energy for nominal material compositions. Since hole quantization energy for such large dots is on the level of a dozen of meV (Fig. 4b), the energy difference between the lowest hole state in the QDs and the valence band edge of the InGaAlAs layer is in the range of 60 meV. In the real structure, the band offsets might be slightly different, the band edges will not be rectangular, and the compositions of both the quaternary layer and QDs might differ slightly from the nominal ones due to the calibration of the MBE growth and the P-As intermixing between the adjacent layers. Despite the energetic proximity, direct transfer of holes to the QDs is not possible due to large (almost 400 nm of InP) spatial separation. However, there are well-known native defects in InP that might mediate this process [Wil1990, Tem1982, Fan1992, Fan1998, Yu1992]. Exemplarily, a complex of indium vacancies $(V_{In})_2$ is typically located 0.36 eV above the InP valence band maximum, $In_PV_{In}$ complex – 0.4 eV, respectively [Wil1990]. Thermal activation of the carrier transfer process can be related either to the activation of holes from the band edge of the InGaAlAs layer to the defect state or from the defect states to QDs. The exact energy of point defects and their complexes (their location in the energy structure) can be very different depending on the growth conditions [Tem1982, Fan1992]. Therefore, lots of them would fit the activation energies determined from the temperature-dependent photoluminescence in this work or shown previously (values in the range of 0.08 eV to 0.32 eV are reported [Fan1992, Fan1998, Yu1992]). It is beyond the scope of this study to identify the exact defects responsible for the described transfer process, and we focus rather on its optical fingerprints and the following implications for the excitation of the QDs. The required energy difference can be supplied by acoustic phonons present in the material even at low temperatures (in our case, 10 K), especially in the case of non-resonant excitation, as they are the main channel of dissipation of the excess energy. The variation of the experimentally determined activation energies reflects the change in the required energy for the transfer process related to the occupation of the QD states by direct optical excitation. The increase in the activation energy for the supply process with increasing excitation power suggests that the defect level in InP has lower energy than the QD states. The observed trend supports this scenario, because when the excitation power is increased, more QD states are occupied directly, and there is a lower number of final states available for the carrier transfer process. The increase in the activation energy corresponds well with the few meV separation of hole states in the QDs (Fig. 4b). Therefore, the temperature at which the maximal PL intensity is achieved reflects the ratio between the direct and InGaAlAs mediated occupation of the QD states. It is correlated with the observed amplitude of the PL intensity increase. Activation energies for PL quenching at low and medium excitation correspond to the escape of electrons into the wetting layer or further into the InP barrier (Fig. 4a)). For the highest excitation power, the activation energy is much lower and in the range of energy difference between the QD hole states and the native InP defect states. This suggests that at this excitation, InP defect-mediated activation to the InGaAlAs layer becomes important because the separation of the QD energy levels is on a much smaller scale - single to dozen meV.

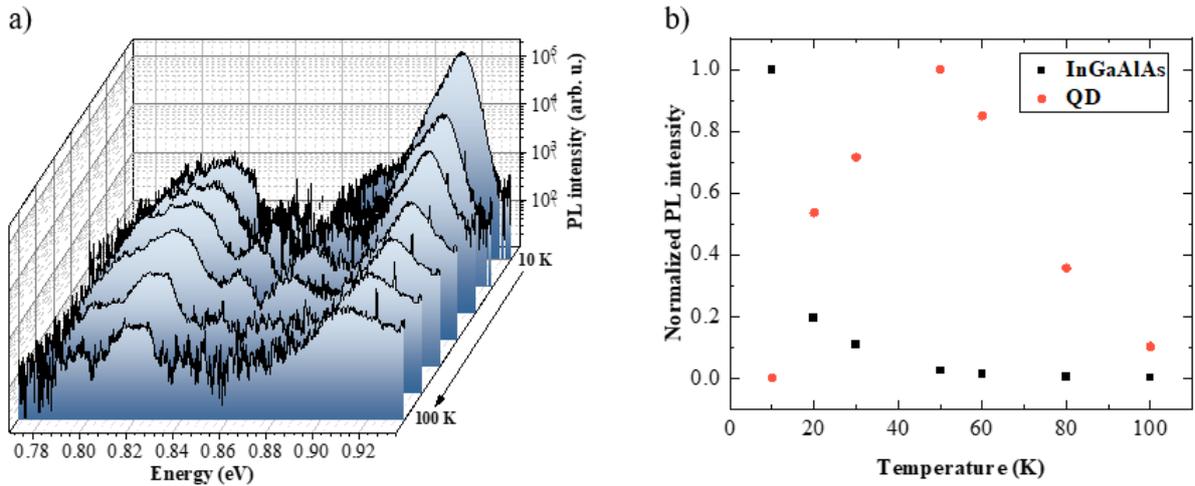

**Figure 5.** a) Low excitation (P = 10 μW) PL spectra measured in the temperature range of (10 – 100) K; low energy emission band corresponds to QD ensemble and high energy emission to InGaAlAs layer; b) normalized emission intensity of QD ensemble (red dots) and InGaAlAs layer (black squares); in both cases, spectrally-integrated emission intensity is considered.

In the results presented so far, only the emission from the QDs was analyzed. In Fig. 5, emission from the QD ensemble and the InGaAlAs layer is presented in the temperature ranges for which the increase in emission intensity was earlier observed for the QD ensemble. A significant change in relative intensities of emission from the InGaAlAs layer and QDs is seen as a function of temperature. At low temperatures, mainly emission from InGaAlAs is observed, and with increasing temperature, emission from the QD ensemble becomes apparent at the expense of decreasing emission intensity centered at 0.914 eV. At even higher temperatures (not shown here), both emissions start to quench. To emphasize that the changes of intensities are correlated, the normalized emission intensity for the QD and InGaAlAs layer is presented in Fig. 5b. Each of the temperature dependences is normalized independently to its maximum intensity, which corresponds to 10 K and 50 K, for the InGaAlAs and QDs, respectively. The relative intensities vary as a function of temperature in an anti-correlated fashion up to 50 K, which supports the scenario of thermally activated carrier transfer from the InGaAlAs layer to the dots. The reversal of the trend for higher temperatures may result from more efficient carrier escape channels in the QDs, activated in this temperature range, due to lower carrier localization energy compared to the InGaAlAs layers.

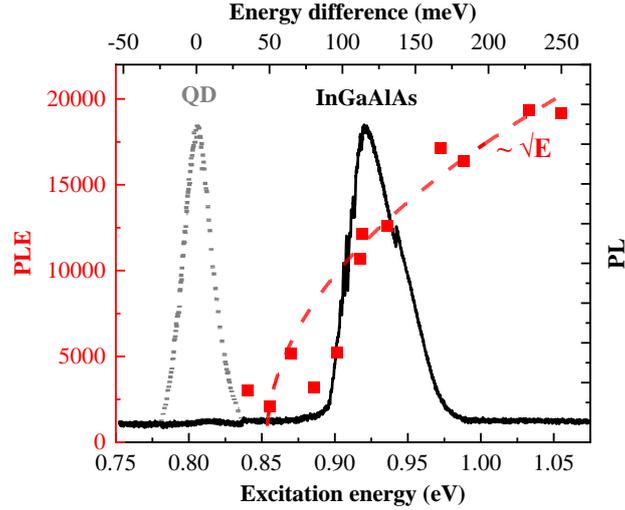

**Figure 6.** Photoluminescence excitation (PLE) signal integrated over the whole spectral range of QD emission (red squares, left axis) overlaid with PL spectra of InGaAlAs (solid black line) and emission from QD ensemble (dashed grey line) measured under non-resonant excitation. The dashed red line represents fit to experimental PLE data with square root energy dependence characteristic for bulk material following the 3D density of states. Experimental conditions: temperature of 5 K, average excitation power of pulsed excitation – 8 μW.

To directly confirm the carrier transfer indicated above, a PLE experiment was conducted, and the results are presented in Fig. 6. The emission from QDs rapidly decreases for the excitation energy below the band edge of the thick InGaAlAs layer (approx. 4 times), and above it seems to reproduce the square root behavior related to the bulk-like density of states and thus its absorption. It proves that the investigated QDs can be efficiently excited through the InGaAlAs layer, explaining the origin of carrier transfer observed in the temperature-dependent PL data presented above.

The fact that the QDs' excitation can be mediated by the InGaAlAs DBR layer and native InP defects has important consequences for applications utilizing both the whole QD ensemble (e.g., lasers and optical amplifiers) and single QD states (e.g., non-classical light sources). In the first case, it provides additional carrier supply and carrier escape channel, which will not be eliminated by changes in the growth mode to avoid the wetting layer formation [Zue2012]. On the other hand, the quaternary material can be engineered in a broad range by changing its composition. However, one has to keep in mind that it will influence the refractive index contrast of DBR layers [Gra2010, Mon1992]. Without optimization, this transfer process will rather degrade the modulation properties of a laser structure.

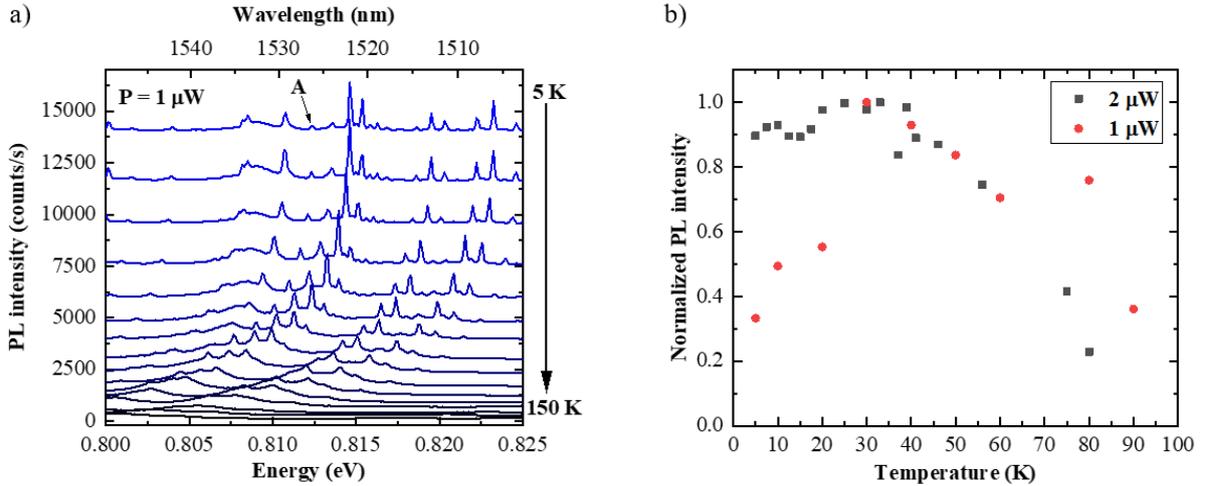

**Figure 7.** a) Temperature series of PL from single QDs in the temperature range of (5 – 150) K for excitation power of 1 μW (cw excitation). b) Normalized integral intensity (from Gaussian fit to experimental data) for line A from a) as a function of temperature for two excitation powers: close to saturation $P_{sat}$ = 2 μW (black squares) and two times lower (red circles) under continuous wave excitation.

Finally, we verify how important is the carrier transfer described above in the case of single QDs. From temperature-dependent measurements on over 15 single QD emission lines, it appears this is not the dominant scenario - only 25% of all the investigated emission lines exhibit an increase in QD emission intensity. However, the low percentage of single QD emission lines showing this behavior might be related to micro-PL measurements being performed on the patterned sample with mesas. In such a structure, a significant part of QDs is near surface charge states that can be present on the rough etched mesa sidewalls. This results in a different local charge environment of a QD compared to the one in a planar sample used for the macro-PL study. Additionally, the carrier transfer requires a specific energy difference between the involved states. In the macro-PL study, a large fraction of the inhomogeneously broadened QD ensemble is probed. Therefore, the probability of finding a QD which fulfills the energy requirements for the carrier transfer is much higher than in the case of single QDs in a mesa structure. The mesa substantially limits the number of QDs probed at the same time - it contains only a small sub-ensemble from the broad distribution of QD properties. When comparing the relative intensity of the emission lines for which the transfer is present (exhibiting an increase in emission intensity with temperature) to other emission lines from the spectrum, it is clear that at low temperature their intensity is much lower and becomes comparable at intermediate temperature range (when the transfer is thermally activated). The temperatures at which the maximal emission intensity is reached for single lines span from 30 K to 70 K. As observed for the QD ensemble, for higher excitation powers, the amplitude of the increase in intensity decreases. Results for an exemplary mesa structure and emission line are presented in Fig. 7. Similarly to the case of the QD ensemble, the maximum PL intensity is reached for a lower temperature in the case of higher excitation. Also, the amplitude of the intensity increase is larger (up to 3 times in the presented example) for lower excitation powers. This shows that the origin of carrier transfer is most probably analogical for single QDs and the ensemble of QDs. However, in the case of single QDs, it is not a dominant effect driving the temperature dependence of emission. For the majority of emission lines (Fig. 7a), a standard behavior – PL quenching – is observed, with two activation energies (on average in the range of single meV and 25 meV) that can be interpreted as an escape of electrons and holes via higher energy states within a single QD, as has already been shown [Smo2021]. The fact that the carrier transfer is not observed for all the emission lines might be related to their different origin – related to different excitonic complexes (charge states) confined in a QD. In the investigated sample majority of the complexes are charged with only a few neutral complexes. The origin of the emission lines for which the transfer effects is visible reflects this distribution – they are mainly charged complexes with only a few neutral ones. As discussed above, only holes can be efficiently transferred to the QDs. Therefore, the carrier transfer prefers a certain charge state. As a result, only the emission from this charge state will exhibit an increase in intensity with temperature. Another possible effect is the inhomogeneous spatial distribution of defects within the InP layer, including the vicinity of individual QDs. Additionally, due to the distribution of the QD energy structure and the character of the transfer process (defect-mediated), it favors only a sub-ensemble of energetically-matched QDs. The inhomogeneous broadening of the QD ensemble, corresponding to 25 meV full width at half maximum, relates to the dot-to-dot differences in the ground state energy. This shows that different conditions need to be fulfilled for efficient carrier transfer to different dots.

The fact that at low excitation and temperature, a QD is mostly occupied via carrier transfer from the InGaAlAs layer might have important consequences for carrier dynamics within the structure. On the one hand, the excitation of QDs via an intermediate state increases the relaxation time and introduces stronger time jitter in the emission process (limiting the photon generation rate and modulation speed). It will also increase the timing jitter between the consecutive emission processes and, therefore, be detrimental to the coherence properties of single and entangled photons (diminished visibility of two-photon interference – degree of their indistinguishability) [Hub2015, Naw2019, Tho2016]. These adverse effects can probably be minimized by resonant excitation schemes, together with optimizing the layer structure to limit the carrier transfer efficiency. On the positive side, it can provide an efficient means for below-barrier excitation of the QDs, combining the advantage of high absorption due to the bulk-like density of states and relatively broad resonant conditions for carrier transfer (related to phonon spectral density at a given temperature).

## Conclusions

In summary, we demonstrate carrier transfer between an InGaAlAs layer forming a bottom DBR and an active medium of an InAs/InP QD structure, mediated by native InP defects. We prove its origin by the temperature dependence of the relative intensity of the dots and InGaAlAs layer emission and by photoluminescence excitation experiment. Described carrier transfer is also present in the case of single QDs. However, it influences only a fraction of optical transitions (25%), as only one type of charge carrier (holes) is involved. Described carrier transfer may have significant consequences for the design and performance of QD-based devices on both a QD ensemble and a single QD level. In particular, its negative influence has to be mitigated. First of all, the emission from the DBR section needs to be filtered out from the detection, which requires filtering wavelengths much closer to the QD emission, if integrated all-fiber configurations are required, imposing additional challenge. The fact that the carriers are not directly relaxing to the QD states, but through the states in the DBR, increases time jitter for occupation of QD states and therefore, emission from the QD. This is detrimental for the time overlap of photons in two photon interference experiment and applications which require Bell-state measurements, e.g., quantum repeater protocols.

## Acknowledgments

This research was funded by the Foundation for Polish Science co-financed by the EU under the ERDF by project entitled „Quantum dot-based indistinguishable and entangled photon sources at telecom wavelengths" carried out within the HOMING program. This work was also financially supported by the BMBF Project (QR.X) and DFG projects (DeLiCom, and Heisenberg grant-BE 5778/4-1). Calculations have been carried out using resources provided by Wroclaw Centre for Networking and Supercomputing (http://wcss.pl), Grant No. 203. We are grateful to Krzysztof Gawarecki for sharing his $\bm{k}\cdot\bm{p}$ method implementation. We also acknowledge Andrei Kors for his assistance in the MBE growth process, Kerstin Fuchs and Dirk Albert for technical support.

**Data availability statement**

The data that supports the findings of this study are available from the corresponding author upon reasonable request.